# Show me the numbers! - Student-facing Interventions in Adaptive Learning Environments for German Spelling

Nathalie Rzepka[1], Katharina Simbeck[1], Hans-Georg Müller[2], Marlene Bültemann[1], Niels Pinkwart[3]

**Abstract:** Since adaptive learning comes in many shapes and sizes, it is crucial to find out which adaptions can be meaningful for which areas of learning. Our work presents the result of an experiment conducted on an online platform for the acquisition of German spelling skills. We compared the traditional online learning platform to three different adaptive versions of the platform that implement machine learning-based student-facing interventions that show the personalized solution probability. We evaluate the different interventions with regards to the error rate, the number of early dropouts, and the users' competency. Our results show that the number of mistakes decreased in comparison to the control group. Additionally, an increasing number of dropouts was found. We did not find any significant effects on the users' competency. We conclude that student-facing adaptive learning environments are effective in improving a person's error rate and should be chosen wisely to have a motivating impact.

**Keywords:** Adaptive Learning, Adaptive Intervention, Learning Analytics.

## 1    Introduction

Adaptive learning environments have been the subject of research for years and are also increasingly used in practice. Being broadly defined and having a large number of possibilities, it is important to examine the effectiveness of different interventions in adaptive learning environments. In this contribution, we specifically investigate adaptive learning interventions for acquisition of spelling skills in German. Our interventions can all be grouped under the term student-facing interventions, i.e., information is displayed to the user in the UI (User Interface). In this case, the student-facing interventions are displays that show the personalized solution probability. Not exclusively, yet especially in the case of student-facing interventions, it should be noted that the influence of the intervention on motivation plays a major role and can influence learning success.

Our article is structured as follows: first, section two explains the theoretical foundations of underlying motivation theory, adaptive learning, and summarizes previous research on student-facing interventions. This is followed by an explanation of the experimental design methodology, the underlying predictive model, the three interventions, and the hypotheses for evaluation. The results and a discussion of the findings follow. In the end, the summary, the limitation of the work as well as implications for further research are given.

[1] HTW Berlin, Treskowallee 8, 10318 Berlin, {rzepka@,simbeck@,marlene.bueltemann@student.}htw-berlin.de
[2] Universität Potsdam, Am neuen Palais 10, 14469 Potsdam, hgmuelle@uni-potsdam.de,
[3] Humboldt-Universität, Unter den Linden 6, 10117 Berlin, pinkwart@hu-berlin.de

## 2 Related Work

### 2.1 Adaptive Learning

Adaptive learning is a concept that aims to optimize the learning success for students through the modification and adaptation of learning content and environments [Wa84]. It is based on the premise that students learn in different ways, which requires different instructional methods, learning paths, or learning characteristics [Wa84]. To meet the students' individual needs, different interventions can be used [Wa84]. Paramythis and Loidl-Reisinger define a learning environment as adaptive "if it is capable of: monitoring the activities of its users; interpreting these on the basis of domain-specific models; inferring user requirements and preferences out of the interpreted activities, appropriately representing these in associated models; and, finally, acting upon the available knowledge on its users and the subject matter at hand, to dynamically facilitate the learning process." [PL03].

[PL03] differentiate between various categories of adaptation in learning environments These are adaptive interaction, adaptive course delivery, content discovery and assembly and adaptive collaboration support. Adaptive interaction describes the adaptation to the user interface of an app or a learning environment. Adaptive course delivery adjusts the content of the course or exercise. Content discovery and assembly uses diverse sources to assemble material and adaption collaboration support focuses on learning processes that involve collaboration and communication. In their review, Wong and Li categorized intervention methods into four categories [WL18]. The first category, direct message, describes all interventions in which a student or a tutor is contacted through messaging., For instance, when they are identified as being at risk or to provide them with additional learning resources. Actionable feedback, the second category, describes all interventions that provide insights and dashboards about the users' learning performance, as well as recommendations to improve ones' learning progress. Categorization of students summarizes interventions that are grouping students based on learning analytics results, such as at-risk predictions. The last category, course redesign aims to adapt the content of a course to the users' need.

One intervention in the category of actionable feedback is student-facing intervention.

## 2.2 Student-facing interventions

One possibility to implement an intervention in an adaptive learning environment may be to show the student his or her performance data [WL18]. In their review, Bodily and Verbert review student-facing learning analytics reporting systems, that directly show students' performance data [BV17]. With 29% and 27% the functionalities "enhanced visualization" and "data mining recommender system" were the most prevalent systems. Enhanced visualization was defined here as the visualization of student data including a class comparison or interactivity feature [BV17]. In their review, they found 14 articles that measured the effects of student-facing reports on student achievements. Of these, eight articles showed significant improvement in student achievements [AP12, CCW08, De14, HHWH09, KJP16, SW14, VKIB13, Wa08] while five had no significant results [DWF15, GB14, ORH15, PJ15, SBP14]. One study had positive and negative effects, depending on the visualization [BHGJ16].

Arnold and Pistilli's research showed how information on students' performance prediction provided directly back to students can improve student retention [AP12]. They proposed an early intervention in which a performance prediction is generated, and the results are sent to the students via e-mail. Additionally, the e-mail contains a traffic light signal indicating how well the student is progressing. Results showed higher retention rates in all years, with higher retention rates the earlier students were exposed to the system [AP12].

Kim et al. validated the impact of a learning analytics dashboard that displays online behavior patterns in college students [KJP16]. They found that student's scores improved in the intervention group compared to the control group. Furthermore, they analyzed the usage frequency and the satisfaction with the dashboard and found that the frequency did not have a significant impact. However, the satisfaction with the dashboard is highest among students who open it infrequently. Users with a high academic score, in contrast, are less satisfied with the dashboard. An earlier study with the same dashboard and fewer participants showed no significant improvement in learning achievement [PJ15].

Chen et al. were able to improve academic performance, task completion rates, and learning goal achievement rates in an experiment that used a ubiquitous learning environment that implemented learning status awareness, schedule reminders, and mentor recommendations [CCW08]. An example with a huge sample size of 50,000 students is the research of [De14]. Here, a course recommendation system is evaluated, and improved the passing grade rate of its users is compared to students who did not use the system [De14]. The results are particularly high for low-income and minority students. Furthermore, Huang et al. evaluated a recommender system based on the Markov chain model to provide learning paths for students [HHWH09]. Users in the treatment group outperformed users in the control group in terms of knowledge acquisition and integration.

The knowledge dashboard in Sauls and Wuttkes' study provided learner insights into the number and scores of questions and tests, strengths and weaknesses, as well as the status of the learning goals [SW14]. Users of the system had higher average grades and lower failure rates [SW14].

Another possibility to intervene in digital learning environments is to display adaptive feedback. Van der Kleij et al. conducted a meta-analysis to compare the effectiveness of different feedback methods on learning outcomes in computer-based learning [vFE15]. They distinguished between three forms of feedback: correctness of the answer, providing the correct answer, and elaborate feedback, which for example provides an explanation. They found that elaborate feedback was more effective than the other two forms of feedback [vFE15]. The results showed that the effects of elaborate feedback on higher-order learning outcomes were greater than on lower-order learning outcomes [vFE15].

While there is already much research about different kinds of student-facing interventions, research lacks student-facing interventions in online language learning. This paper, therefore, presents an online-controlled experiment that compares the effectiveness of adaptive learning in three different student-facing interventions on a German spelling learning platform to the control group. For this purpose, we transformed a learning platform into an adaptive learning platform and implemented a machine learning-based prediction model on which the interventions are based. We evaluate the interventions based on three different research questions:

RQ 1: How effective are student-facing interventions on an online spelling platform in terms of the error rates?

RQ 2: How effective are student-facing interventions on an online spelling platform concerning the number of early dropouts?

RQ 3: How effective are student-facing interventions on an online spelling platform concerning the users' competency?

RQ 4: How do effects differ with regards to different implementations of student-facing interventions?

## 3  Methodology

The platform Orthografietrainer.net is a platform to support the acquisition of German spelling and grammar skills. It is a free, web-based platform that currently has more than one million registered users from Germany, Austria, and Switzerland. The platform offers spelling exercises on various orthographic areas, e.g., comma formation, capitalization, hyphenation, and sounds and letters. The exercises are suitable for students from fifth grade onwards, as well as adults, university students, or older school children are among the users. Typically, the platform is used in blended classroom scenarios. Thus, the students are registered by their teachers and receive the login credentials during class. The teacher can assign homework their students, which is then displayed to the students as pending. These exercise sets usually consist of ten exercise sentences with increasing difficulty. After each sentence, the user receives immediate feedback as to whether the solution was correct or not. If not, the task is repeated, and the user must solve more similar sentences before moving on to the more difficult sentences. At the end, the user receives an overview of his or her results. A teacher is provided with a dashboard that shows the progress and results of all the students.

We conducted an online-controlled experiment that was carried out from the 21$^{st}$ of June to the 31$^{st}$ of October in 2022. During this time, all users in the student user group who performed capitalization tasks were randomly assigned to the control group or one of three intervention groups. All three intervention groups adapt to the user based on the prediction of the users' performance. Tab. 1 shows the distribution of users across the different groups. The experiment was pre-registered at the OSF[4] and as an architectural setup described in [Rz22].

| Intervention group | Number of users | Number of sessions | Number of answered sentences |
|---|---|---|---|
| Control | 2,447 | 8,049 | 225,426 |
| Intervention 1 | 1,835 | 5,950 | 148,625 |
| Intervention 2 | 1,929 | 6,222 | 159,677 |
| Intervention 3 | 1,910 | 6,072 | 153,658 |

Tab. 1: Experiment metrics per intervention group

The adaptive learning environment includes a prediction model, which predicts the probability of correctly processing the next sentence by the user. It was trained using a dataset from the Orthografietrainer.net platform and contained 200.000 records of capitalization performed by students from grades five through twelve. After feature engineering, the dataset contains 1078 features: 6 features that refer to demographic data of the user, such as gender or grade level; 17 features that describe the upcoming sentence, including its difficulty as well as information on former attempts of the user to solve this sentence. The other 1055 features represent the users' learning history on the platform. Each feature indicates for an exercise sentence whether the user has already processed this sentence and if so, whether his or her attempt was correct or not. Using this dataset, we trained a decision tree classifier and were able to predict a user's probability of correctly solving the sentence with an accuracy of 97,04% (Recall: 96,31%; Precision: 97,75%). The decision tree model was chosen over other model implementations because of inferior accuracy or performance.

In our experiment, we compared the control group and three interventions (Tab. 2). The control group does not receive any adaptive interventions on top of the established platform. Interventions 1 and 2 are student-facing interventions where users are shown their prediction results. In Intervention 1, the prediction result is shown verbally (see Tab. 3 and Fig. 1, left), in intervention 2 it is shown as a percentage (see Fig. 1, right). Intervention 3 does not show the prediction results. Instead, for users whose prediction result is below 50%, the suitable spelling rule is displayed (see Fig. 2).

---
[4] 10.17605/OSF.IO/3R5Y7

| Group | Description |
| --- | --- |
| Control | No adaptive interventions. |
| Intervention 1 | Prediction results are shown to the user verbally. |
| Intervention 2 | Prediction results are shown to the user as a percentage. |
| Intervention 3 | If the prediction is <50% to solve that sentence correctly, the orthographical rule is shown to the user. |

Tab. 2: Overview adaptive interventions.

| P | Verbal display |
| --- | --- |
| P<0.2 | Attention! This sentence is especially difficult for you. |
| P<0.4 | Beware! This sentence is difficult for you. |
| P<0.6 | Think carefully! This sentence is moderately difficult for you. |
| P<0.8 | Relax! This sentence is easy for you. |
| P>0.8 | No problem! This sentence is especially easy for you. |

Tab. 3: Verbal prediction results (translated from German into English).

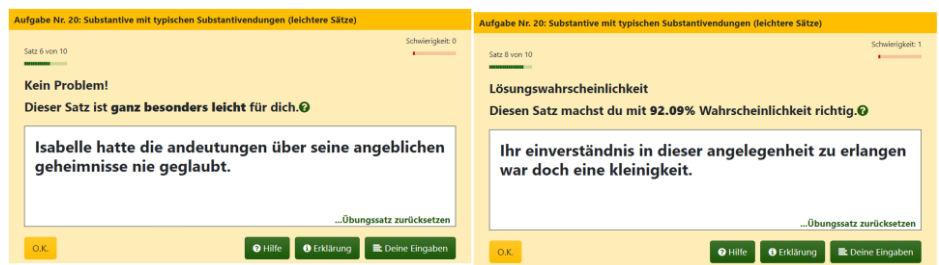

Fig. 1: Example of intervention 1 (left): Verbal prediction result ("No problem! This sentence is especially easy for you."). Example of intervention 2 (right). Display of prediction result as a percentage ("You are 92.09% likely to get this sentence right.")

Fig. 2: Example of intervention 3: Display of spelling rule in the grey box. Message: "Attention! See the spelling rule of this exercise on the right-hand side."

**Evaluation.**

The effects of adaptive learning interventions are evaluated using three hypotheses. In hypothesis 1 we expect a change in the relative number of incorrect answers. We calculate the relative number of incorrect answers by dividing all incorrect answers per user by all given answers per user. In hypothesis 2, we expect a change in the number of early dropouts. As stated before, each exercise set consists of ten sentences. A session is defined as dropped out if there are more than 45 minutes between two processed sentences. We set 45 minutes as the threshold because, after 45 minutes of inactivity, the user is logged out automatically. If the user leaves the platform before he or she finished the whole exercise set, it is possible to continue later in time. However, this will count as a new session. In hypothesis 3 we expect a change in the users' competency. We calculate the competency of the users with the Rasch model, an implementation of the item response theory (IRT). Here, not only the users' responses are included, but also the item difficulty. In the statistical analysis, we first test the assumptions of homogeneity of variance and normal distribution. Since the assumptions were not met in all three hypotheses, we continue with non-parametric tests. For each hypothesis, we perform a Kruskal-Wallis-Test. In case of significant results, we proceed with a Wilcoxon-Mann-Whitney-Test. As multiple tests are performed, we carry out a Bonferroni correction. After that, the level of significance is set at 0.017. Effect sizes are calculated with Cliff's Delta [Cl93]. We set the threshold for a small effect at .11 and at .28 for a medium effect, as in [VDV00].

## 4 Results

### 4.1 H1 – Incorrect answers

With regards to the relative number of incorrect answers we found significantly fewer mistakes in all interventions compared to the control group. Here, interventions 1 and 2 result in an effect size of 0.11 and 0.12, while intervention 3 produces an effect size of 0.09 and is therefore negligible (Tab. 4). The mean percentage of errors in the control group is 16.98% and which is higher as the mean percentage of errors in all other groups (Intv 1: 13.91%, Intv 2: 14.32%, Intv 3: 14.77%).

| p-value | Control | Intv 1 | Intv 2 | Effect sizes | Control | Intv 1 | Intv 2 |
|---|---|---|---|---|---|---|---|
| Intv 1 | **3.99e-12** | | | Intv 1 | **0.1237** | | |
| Intv 2 | **9.43e-12** | 0.8788 | | Intv 2 | **0.1198** | -0.0029 | |
| Intv 3 | **2.91e-08** | 0.1846 | 0.2588 | Intv 3 | 0.0978 | -0.0250 | -0.0210 |

Tab. 4: Results of H1- incorrect answers. Left: p-value, significant results in bold font. Right: effect sizes, results with small effect in bold font; medium effects in italic font.

Interestingly, the results for intervention group 3 differ for boys and girls. While boys only have a median of relative mistakes of 16.05% (effect size: 0.07), for girls it results in 13.48% (effect size: 0.12). Hence, girls appear to benefit most from the display of the rule, while for boys, this is the least effective intervention.

### 4.2  H2 – Dropout

An analysis of the number of dropouts (H2) showed significantly higher dropouts in comparison to the control group in all intervention groups. The highest mean number of dropouts per user was found in intervention 2 with 15.51% (control group: 12.33%). Intervention 1 had a mean number of dropouts of 14.06%; intervention 3 of 14.25%. However, the effect sizes are negligible.

| p-value | Control | Intv 1 | Intv 2 | Effect sizes | Control | Intv 1 | Intv 2 |
|---|---|---|---|---|---|---|---|
| Intv 1 | **0.0027** | | | Intv 1 | -0.0173 | | |
| Intv 2 | **4.70e-08** | 0.0251 | | Intv 2 | -0.0317 | -0.0144 | |
| Intv 3 | **0.0009** | 0.7790 | 0.0491 | Intv 3 | -0.0191 | -0.0018 | -0.0126 |

Tab. 5: Results of H2- Number of dropouts. Left: p-value, significant results in bold font. Right: effect sizes, results with small effect in bold font; medium effects in italic font.

### 4.3  H3 – User competency

The last hypothesis H3 compared the mean competency that is calculated by the Rasch model. As seen in Tab. 6 differences between groups are not significant.

| p-value | Control | Intv 1 | Intv 2 | Effect sizes | Control | Intv 1 | Intv 2 |
|---|---|---|---|---|---|---|---|
| Intv 1 | 0.4901 | | | Intv 1 | 0.0124 | | |
| Intv 2 | 0.3242 | 0.8124 | | Intv 2 | 0.0175 | 0.0045 | |
| Intv 3 | 0.0569 | 0.2761 | 0.3751 | Intv 3 | 0.0340 | 0.0208 | 0.0167 |

Tab. 6: Results of H3: User competency. Left: p-value, significant results in bold font. Right: effect sizes, results with small effect in bold font; medium effects in italic font.

## 5 Discussion

In our article, we present the results of an online-controlled experiment that compares a traditional online learning platform for German spelling skills to three adaptive versions of the platform implementing student-facing interventions. We evaluate the experiment results concerning the error rate, the number of dropouts, and the users' competency.

We found that error rates decreased significantly for all users in the intervention groups in comparison to the users that were in the control group. Hence, the student-facing interventions had positive effects on that metric. Users in intervention 3 received the spelling rule as a hint if the solving probability is below 50%. Here, we found that this is particularly effective for girls, but the least effective intervention for boys. One could interpret that it is more likely for girls to read the spelling rule carefully and use it to solve the exercise while boys are not using the additional information that is displayed.

Concerning the dropout rate, we found that dropouts increased in all intervention groups in comparison to the control group. This could mean, that users get demotivated when they receive low solving probabilities and thus leave the session. Furthermore, leaving a session after receiving a very good prediction could have two reasons: first, users might get bored or think that they don't need to practice anymore. Secondly, users could become demotivated if they receive a very high prediction score and still fail. This also goes hand in hand with the theoretical foundations of motivation theory. In flow theory, users are most persistent when the tasks are neither too easy nor too difficult. It should be discussed if dropping out can only be seen as a negative consequence. It is clear that a dropout can be out of frustration or demotivation. However, there are many other reasons why a user does not to finish the exercise set that are neither positive nor negative. This could be because the school lesson is over, or homework is interrupted for private reasons. Furthermore, a dropout can be a positive effect of student-facing interventions. For example, if the student feels that it is too difficult and the student-facing intervention shows him or her the same, then the student's emotions are validated, and one could decide to do some simpler sets first or to take a break. From a pedagogical point of view, that would be a positive effect of student-facing interventions that leads to a dropout.

We did not find any significant intervention groups with regard to the users' competence. As spelling competence does not develop quickly, it seems that the experimental period was not long enough to find significant differences. Furthermore, we did not do a pre-and post-test of the users to detect changes in time. Instead, we used all exercises that were solved during the experimental period to calculate the person's ability estimates.

## 6 Conclusion

Our article presents the results of an online-controlled experiment that was carried out on an online platform for the acquisition of German spelling skills from June to October 2022. For that, we implemented a machine learning-based prediction model that calculated the personalized solving probability for a user and an exercise at hand. We further implemented three different student-facing interventions that all used the prediction results. Our results showed that all three interventions led to a decreasing error rate for the users in comparison to the control group. As dropout numbers increased, we discussed the meaning of a dropout and found a connection between dropouts and the prediction received by the user. The calculation of the users' competencies did not show significant results. In summary, we found that student-facing machine learning-based interventions lead to fewer errors in German spelling learning environments. However, it can also demotivate users leading to more dropouts. An experimental period of four months is not enough to conclude the long-term development of competencies.

Our study comes with limitations that are discussed in the following. First, our experimental period did not cover a whole school year, but only four months. Moreover, it was conducted during summertime, including the German summer school break. Furthermore, concerning the competency calculation, we did not conduct pre-and post-tests to compare the users' developments.

Further research should therefore include standardized pre- and post-test to be able to conclude the users' competency. Additionally, one could also focus on more orthographical areas and other languages. Especially a comparison of the effects on first-language and second-language learning would be of interest.

## Bibliography

AP12    Arnold, K. E.; Pistilli, M. D.: Course signals at Purdue. In (Dawson, S. Ed.): Proceedings of the 2nd International Conference on Learning Analytics and Knowledge. ACM, New York, NY, p. 267, 2012.

Be16    Beheshitha, S. S. et al.: The role of achievement goal orientations when studying effect of learning analytics visualizations. In (Gaševic, D. Ed.): Proceedings of the Sixth International Conference on Learning Analytics & Knowledge. ACM, New York, NY, pp. 54–63, 2016.

BV17   Bodily, R.; Verbert, K.: Review of Research on Student-Facing Learning Analytics Dashboards and Educational Recommender Systems. IEEE Transactions on Learning Technologies 4/10, pp. 405–418, 2017.

CCW08  Chen, G.; Chang, C.; Wang, C.: Ubiquitous learning website: Scaffold learners by mobile devices with information-aware techniques. Computers & Education 1/50, pp. 77–90, 2008.

Cl93   Cliff, N.: Dominance statistics: Ordinal analyses to answer ordinal questions. Psychological Bulletin 3/114, pp. 494–509, 1993.

De14   Denley, T.: How Predictive Analytics and Choice Architecture Can Improve Student Success. Research & Practice in Assessment 9, pp. 61–69, 2014.

DWF15  Dodge, B.; Whitmer, J.; Frazee, J. P.: Improving undergraduate student achievement in large blended courses through data-driven interventions. In (Baron, J. et al. Eds.): Proceedings of the Fifth International Conference on Learning Analytics And Knowledge. ACM, New York, NY, USA, pp. 412–413, 2015.

GB14   Grann, J.; Bushway, D.: Competency map. In (Pistilli, M. et al. Eds.): Proceedins of the Fourth International Conference on Learning Analytics And Knowledge - LAK '14. ACM Press, New York, New York, USA, pp. 168–172, 2014.

Hu09   Huang, Y.-M. et al.: A Markov-based Recommendation Model for Exploring the Transfer of Learning on the Web. Journal of Educational Technology & Society 2/12, pp. 144–162, 2009.

KJP16  Kim, J.; Jo, I.-H.; Park, Y.: Effects of learning analytics dashboard: analyzing the relations among dashboard utilization, satisfaction, and learning achievement. Asia Pacific Education Review 1/17, pp. 13–24, 2016.

Ot15   Ott, C. et al.: Illustrating performance indicators and course characteristics to support students' self-regulated learning in CS1. Computer Science Education 2/25, pp. 174–198, 2015.

PJ15   Park, Y.; Jo, I.: Development of the learning analytics dashboard to support students' learning performance. Journal of Universal Computer Science 1/21, p. 110, 2015.

PL03   Paramythis, A.; Loidl-Reisinger, S.: Adaptive learning environments and e-learning standards. In (Williams, R. Ed.): 2nd European Conference on e-Learning. Glasgow Caledonian University, Glasgow, 6-7 November 2003. Academics Conferences International, Reading, UK, 2003.

Rz22 Rzepka, N. et al.: An Online Controlled Experiment Design to Support the Transformation of Digital Learning towards Adaptive Learning Platforms: Proceedings of the 14th International Conference on Computer Supported Education. SCITEPRESS - Science and Technology Publications, 2022.

SBP14 Santos, O. C.; Boticario, J. G.; Pérez-Marín, D.: Extending web-based educational systems with personalised support through User Centred Designed recommendations along the e-learning life cycle. Science of Computer Programming 88, pp. 92–109, 2014.

SW14 Saul, C.; Wuttke, H. D.: Turning Learners into Effective Better Learners: The Use of the askMe! System for Learning Analytics, 2014.

VDV00 Vargha, A.; Delaney, H. D.; Vargha, A.: A Critique and Improvement of the "CL" Common Language Effect Size Statistics of McGraw and Wong. Journal of Educational and Behavioral Statistics 2/25, p. 101, 2000.

Ve13 Vesin, B. et al.: Applying Recommender Systems and Adaptive Hypermedia for e-Learning Personalizatio. COMPUTING AND INFORMATICS 3/32, pp. 629–659, 2013.

vFE15 van der Kleij, F. M.; Feskens, R. C. W.; Eggen, T. J. H. M.: Effects of Feedback in a Computer-Based Learning Environment on Students' Learning Outcomes. Review of Educational Research 4/85, pp. 475–511, 2015.

Wa08 Wang, F.: Content recommendation based on education-contextualized browsing events for web-based personalized learning, 2008.

Wa84 Wang, M.: The Adaptive Learning Environments Model: Design, Implementation, and Effects, 1984.

WL18 Wong, B. T.; Li, K. C.: Learning Analytics Intervention: A Review of Case Studies. In (Wang, F. L. Ed.): 2018 International Symposium on Educational Technology. ISET 2018  31 July-2 August 2018, Osaka, Japan  proceedings. IEEE, Piscataway, NJ, pp. 178–182, 2018.